\documentclass[aps,preprint,showpacs,superscriptaddress,groupedaddress]{revtex4}  
\usepackage{graphicx}  
\usepackage{dcolumn}   
\usepackage{bm}        
\usepackage{amssymb}   
\usepackage{amsmath}

\begin{document}


\title{On the Angular Momentum of Rockets, Balloons, and Other Variable Mass Systems}
%
%
\author{A.~Nanjangud} \affiliation{Department of Mechanical and Aerospace Engineering,~The
University of California,~Davis, One Shields Avenue, Davis, CA, 95616} \affiliation{Jet
  Propulsion Laboratory, California Institute of Technology, Pasadena, California 91109}
\author{F.O.~Eke$^1$}
\vskip 0.25cm
\date{\today}

\begin{abstract}
Variable mass systems are a classic example of open systems in classical
mechanics. The reaction forces due to mass variation propel ships,
balloons, and rockets. Unlike free constant mass systems, the angular
momentum of these systems is not of constant magnitude due to the change
in mass. In this paper, we show that the angular momentum vector for such
a system has a fixed direction in space and, thus, is partially conserved
for both rigid and flexible, torque-free, variable mass systems. A
potential use of this result is that it provides a suitable stationary
reference frame against which the orientation of variable mass system could
be measured.
\end{abstract}

\maketitle
Mass-varying systems have a longstanding history in classical mechanics.
According to \v{S}\'{i}ma and Podolsk\'{y} \cite{sima}, Buquoy first
formulated the equations of motion for a generic variable mass system and
also offered some context to its application by examining the vertical
motion of an extensible flexible fiber. However, Mikhailov \cite{mikhailov}
considers Bernoulli's idea of jet-propelled ships \cite{bernoulli} as the
first foray into reaction devices and, thus, moving variable mass systems.
Moore \cite{moore}, around the same time as Buquoy, focused on a different
subset of variable mass systems by deriving the classic rocket equation.
Since then, most of the academic and engineering discourses on variable
mass system dynamics have focused primarily on their application to rockets.

The pioneering works of stalwarts such as Tsiolkovsky \cite{tsiolkovsky} and
Goddard \cite{goddard} was on understanding the translational dynamics of
rockets to escape earth's gravitational field. However, a flurry of work
concentrating on the rotational dynamics of rockets and variable mass
systems is seen in the mid-twentieth century. Rosser et al \cite{rosser}
examined the motion of both spinning and non-spinning rockets and are considered
to be the first to present the idea of jet damping, {\sl apud} van der Ha and
Janssens \cite{vanderha}. Forms of the rotational equations that more closely
resemble Euler's rigid body equations are seen in a paper by Ellis and
McArthur \cite{ellis}, and Thomson's \cite{thomsontext} classic textbook, with
the latter addressing the general variable mass system with discrete mass loss.
A more sophisticated model utilizing a control volume approach to account for
continuous mass variation was later presented by Thorpe \cite{thorpe}. This
control volume approach has since gained widespread acceptance \cite{meirovitch,
cornelisse, eke, quadrelli}.  We consider it interesting to note that, despite
interest in variable mass systems beginning in the same era as Euler's study of
rigid body motion, analytical studies into their rotational motion have only
been performed over the last 70 years.

Our letter also concerns the rotational dynamics of variable mass systems,
wherein we prove the conservation of the direction of angular momentum about
the instantaneous mass center. The utility of this result can be understood from
the perspective of some developments on constant mass systems' dynamics. In the
case of free constant mass systems, proving the fixedness of the angular
momentum vector is a trivial outcome, due to its conservation, but has proved
extremely valuable in theory \cite{poinsot} as well as in practice
\cite{bracewell, landon}. Poinsot \cite{poinsot} developed an approach to
visualize the motion of asymmetric rotating rigid bodies without explicitly
solving the governing nonlinear equations of motion. His work is considered one
of the cornerstones in the field of geometric mechanics. Relatively recently,
Poinsot's work was extended to explain the dissipative effects of flexibility
on spinning structures, independently by Bracewell and Garriott
\cite{bracewell}, and Landon and Stewart \cite{landon}. These investigators
showed that flexible bodies (without mass variation) rotate stably about the
axis of maximum moment of inertia but are unstable about the axes of
intermediate and minor moment of inertia, a phenomenon verified by the unstable
flight of the Explorer-1 satellite. The fixed nature of the angular momentum
vector also permits solutions to orientation angles (e.g. Euler angles) of
rotating axisymmetric rigid bodies \cite{likins} and such analytical results
are useful in controlling the motion of bodies. The development of similar
literature on variable mass systems can be extremely useful at a time when
spaceflight is more commonplace but a better understanding of these systems'
angular momenta is necessary to enable such work.

\begin{figure}[h]
  \includegraphics[scale=1.4]{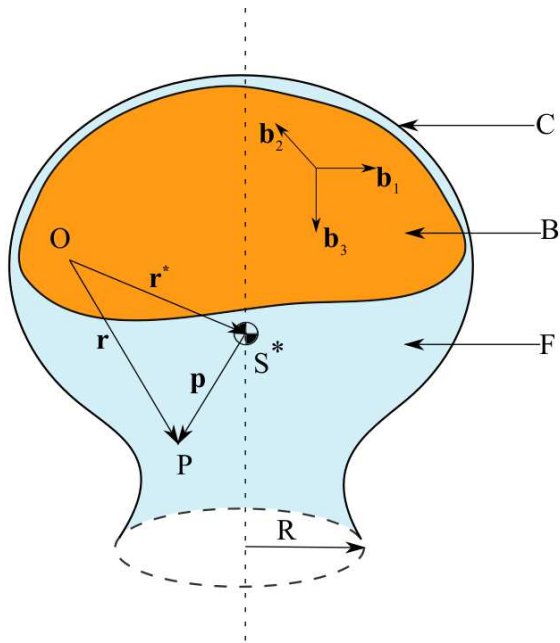}
  \caption{\label{fig:1}General Variable Mass System}
\end{figure}
Figure \ref{fig:1} is that of a system with mass variation comprising a
consumable rigid base $B$ and a fluid phase $F$. A massless shell $C$ of
volume $V_0$ and surface area $S_0$ is attached to $B$. It is assumed that
mass can enter or exit $C$ through the region represented as a dashed ellipse.
The shell and everything within it is considered to be of interest, while any
matter outside of it is not. At any instant, there is a definite set of matter
within the region $C$ which obeys the laws of mechanics. At another instant,
$C$ may contain a different set of matter but it too must obey the laws of
mechanics at that instant. Thus, the angular momentum principle can be applied
to $C$ and its contents to derive the vector equation of attitude motion that
are valid at each instant of time.

At any general instant of time there is a definite set of matter within $C$.
The angular momentum of this constant mass system about its mass center
$S^*$, denoted ${\bf H}^*$, is given by
\begin{equation}
   \label{3.1}
   {\bf H}^* = \int_{V} \rho {\bf p \times v} \, \mathrm{d}V,
\end{equation}
where $V$ is the volume occupied by the contents of the constant mass system at
the instant of interest, $\rho$ is the mass density, $\bf p$ is a position
vector from $S^*$ to an arbitrary particle $P$ within $C$, and $\bf v$ is the
inertial velocity of $P$. The motion of $P$ when observed from $B$  allows the
above angular momentum expression to be reformulated as
\begin{equation}
  \label{3.2}
   {\bf H}^* = \int_{V} \rho {\bf p} \times [{\bf v}^o 
    + {\bf v}_r + {\bm{\omega} \times \bf r}] \, \mathrm{d}V,
\end{equation}
where $\bf r$ is a position vector from $O$ to $P$, ${\bf v}^o$ is the inertial
velocity of $O$, ${\bf v}_r$ is the velocity of $P$ relative to $B$, and
$\bm{\omega}$ is the inertial angular velocity of $B$. Equation (\ref{3.2}) is
then expanded as 
\begin{equation}
    \label{3.3}
    {\bf H}^* = \int_{V} \rho {\bf p} \, \mathrm{d}V \times {\bf v}^o
    + \int_{V} \rho {\bf p} \times {\bf v}_r \, d\mathrm{}V
    + \int_{V} \rho {\bf p} \times ({\bm \omega \times \bf r}) \, \mathrm{d}V,
\end{equation}
The first integral on the right-hand side of Equation (\ref{3.3}) evaluates to
zero by virtue of the definition of a mass center. Further, from Figure
\ref{fig:1}, it is evident that ${\bf r} = {\bf r}^* + {\bf p}$ where
${\bf r}^*$ is the position vector from $O$ to $S^*$ so Equation (\ref{3.3}) can
be rewritten as
\begin{equation}
  \begin{aligned}
    \label{3.4}
    {\bf H}^* = \int_{V} \rho {\bf p} &\times {\bf v}_r \, \mathrm{d}V
    + \int_{V} \rho {\bf p} \times ({\bm{\omega} \times {\bf r}^*}) \, \mathrm{d}V
    + \int_{V} \rho {\bf p} \times ({\bm{\omega} \times {\bf p}}) \, \mathrm{d}V. 
  \end{aligned}
\end{equation}
The second volume integral on the right-hand side of Equation (\ref{3.4})
evaluates to zero, again, by the principle of mass centers. Equation (\ref{3.4})
is now written in a compact form as 
\begin{equation}
    \label{3.5}
    {\bf H}^* = \int_{V} \rho {\bf h} \, \mathrm{d}V,
\end{equation}
where ${\bf h}$ is
\begin{equation}
    \label{3.6}
    {\bf h} \triangleq {\bf p \times[{\bf v}_r + (\bm{\omega} \times p})].
\end{equation}
Equation (\ref{3.5}) now gives the instantaneous angular momentum of a constant
mass system. The angular momentum principle applied to this constant mass system
about its mass center is
\begin{equation}
  \label{3.7}
  {\bf M}^* = \frac{^N\mathrm{D}{\bf H}^*}{\mathrm{D}t},
\end{equation}
where ${\bf M}^*$ is the sum of all moments due to external forces on the
constant mass system, and $\frac{^N\mathrm{D}}{\mathrm{D}t}$ is the material
derivative observed from an inertial frame $N$. In the case of torque-free
motion, ${\bf M}^*= {\bf 0}$ which when used in Equation (\ref{3.7}) gives
\begin{equation}
  \label{3.8}
  {\bf 0} = \frac{^N\mathrm{D}}{\mathrm{D}t}\int_{V} \rho {\bf h} \, \mathrm{d}V.
\end{equation}
Note that, in Equation (\ref{3.8}), ${\bf H}^*$ has been expressed in its
integral form, given by Equation (\ref{3.5}). The above equation tells us that
the angular momentum of the constant mass system is invariant. If we choose to
switch from the inertial reference frame to a reference frame attached to $B$
then Equation (\ref{3.8}) can be rewritten as
\begin{equation}
  \label{3.9}
  {\bf 0} = \frac{^B\mathrm{D}}{\mathrm{D}t}\int_{V} \rho {\bf h} \, \mathrm{d}V
  + \bm{\omega} \times \int_{V} \rho {\bf h} \, \mathrm{d}V.
\end{equation}
In the above form, the two terms on the right hand side of Equation (\ref{3.9})
focus on the constant mass system. Attention can be transferred to the control
volume with fluxing matter via two operations. Firstly, Reynolds' Transport
Theorem  is invoked on the first term on the right-hand side of Equation
(\ref{3.9}). Secondly, noticing that, at the instant for which the above
equation is derived, $V= V_0$. As a result, Equation (\ref{3.9}) becomes
\begin{equation}
  \label{3.10}
  {\bf 0} = \frac{^B\mathrm{d}}{dt}\int_{V_0} \rho {\bf h} \, \mathrm{d}V
  + \int_{S_0} \rho {\bf h} ({\bf v}_r\cdot {\hat{\bf n}}) \, \mathrm{d}S
  + \bm{\omega} \times \int_{V} \rho {\bf h} \, \mathrm{d}V.
\end{equation}
In the above equation, $\hat {\bf n}$ is an outwardly directed unit normal from
a surface of $C$ through which mass enters and/or exits. If ${\bf v}_r \cdot
{\hat{\bf n}} = u$, where $u$ is a general scalar variable, Equation
(\ref{3.10}) can be rewritten as
\begin{equation}
  \label{3.11}
  {\bf 0} = \frac{^B\mathrm{d}{\bf H}^*_0}{\mathrm{d}t}
  + \int_{S_0} \rho {\bf h} u \, \mathrm{d}S
  + \bm{\omega} \times {\bf H}_0^*,
\end{equation}
where ${\bf H}^*_0$ is the angular momentum of the variable mass system and is
\begin{equation}
    \label{3.12}
    {\bf H}^*_0 = \int_{V_0} \rho {\bf h} \, \mathrm{d}V.
\end{equation}
Since $V = V_0$ at a particular instant, ${\bf H}^*$ and ${\bf H}_0^*$ are
identical but their time derivatives are generally not identical since their
evolution in time is associated with changing sets of matter. Since our interest
is in understanding the behaviour of the variable mass system's angular momentum
from an inertial frame, we revert the time derivative in Equation (\ref{3.11})
to $N$
\begin{equation}
  \label{3.13}
  {\bf 0} = \frac{^N\mathrm{d}{\bf H}^*_0}{\mathrm{d}t}
  + \int_{S_0} \rho {\bf h} u \, \mathrm{d}S.
\end{equation}
Any vector can be expressed as a combination of a scalar and a unit vector
directed along the vector itself. So, $\bf h$ is rewritten as ${\bf h} =
h{\hat {\bf n}_h}$, where ${\hat {\bf n}_h}$ is a unit vector directed along
$\bf h$ whose magnitude is $h$. As a result, Equation (\ref{3.12}) can be
written as
\begin{equation}
  \label{3.14}
  {\bf H}^*_o = \bigg(\int_{V_o} \rho h \, \mathrm{d}V \bigg){\hat {\bf n}_h}
\end{equation}
and Equation (\ref{3.13}) as
\begin{equation}
  \label{3.15}
  \frac{^N\mathrm{d}{\bf H}^*_o}{\mathrm{d}t}
  = \bigg(-\int_{S_o} \rho h u \,\mathrm{d}S\bigg) {\hat {\bf n}_h}
\end{equation}
Equation (\ref{3.14}) asserts that ${\bf H}^*_o$ is not of constant magnitude
while Equations (\ref{3.14}) and (\ref{3.15}) assert that it is always
directed along the $\hat{\bf n}_h$ vector, which it will now be proved is an
inertially fixed vector. 

Let $\hat {\bf n}_f$ and $\hat {\bf n}_g$ be two unit vectors which form a
dextral set with $\hat{\bf n}_h$, then $\hat {\bf n}_f \times \hat {\bf n}_g
= \hat {\bf n}_h$ and so on. This dextral set of unit vectors are attached to an
imaginary reference frame $Q$ whose inertial angular velocity is expressed as
\begin{equation}
  \label{3.16}
  {\bm \omega}^Q = \Omega_1 \hat{\bf n}_f+ \Omega_2 \hat{\bf n}_g+\Omega_3 \hat{\bf n}_h.
\end{equation}
The time rate of change of $\hat {\bf n}_h$ in the inertial frame is
\begin{equation}
  \label{3.17}
  \frac{^N{\mathrm d}\hat {\bf n}_h}{\mathrm dt} = \frac{^Q\mathrm{d}\hat {\bf n}_h}{\mathrm dt}
  + {\bm \omega}^Q \times \hat {\bf n}_h
\end{equation}
where the first term on the right-hand side of Equation (\ref{3.17}) evaluates
to zero since $\hat {\bf n}_h$ is fixed in $Q$. Then, substituting for
$\bm \omega^Q$ from Equation (\ref{3.16}) in Equation (\ref{3.17}) gives
\begin{equation}
  \label{3.18}
  \frac{^N {\mathrm d}\hat {\bf n}_h}{\mathrm dt} = \Omega_2 \hat{\bf n}_f
  - \Omega_1 \hat{\bf n}_g.
\end{equation}

Further, Equation (\ref{3.15}) is rewritten as
\begin{equation}
  \label{3.19}
  \frac{^N\mathrm{d}}{\mathrm{d}t}\int_{V_o} \rho h \, \mathrm{d}V \hat {\bf n}_h
  = \bigg(-\int_{S_o} \rho u h \, \mathrm{d}S\bigg) \hat {\bf n}_h
\end{equation}
or
\begin{equation}
    \label{3.20}
    \frac{\mathrm d}{\mathrm dt}\bigg( \int_{V_o} \rho h \, {\mathrm d}V \bigg)
    \hat {\bf n}_h
    + \bigg(\int_{V_o} \rho h \, \mathrm dV\bigg) \frac{^N{\mathrm d}
    {\hat {\bf n}_h}}{\mathrm dt}\\
    = \bigg( -\int_{S_o} \rho u h \, \mathrm dS \bigg)\hat {\bf n}_h.
\end{equation}
The result from Equation (\ref{3.18}) is substituted in Equation (\ref{3.20})
to give
\begin{equation}
  \begin{aligned}
  \label{3.21}
  \frac{\mathrm d}{\mathrm dt}
  \bigg( \int_{V_o} &\rho h \, {\mathrm d}V \bigg) \hat {\bf n}_h
  + \bigg( \int_{V_o} \rho h \, {\mathrm d}V \bigg)
  (\Omega_2 \hat{\bf n}_f - \Omega_1 \hat{\bf n}_g)\\
  &= \bigg( -\int_{S_o} \rho u h \, {\mathrm d}S \bigg)\hat {\bf n}_h.
  \end{aligned}
\end{equation}
The above equation, when rewritten in component form, leads to $\Omega_1 =
\Omega_2 = 0$. Using these values for $\Omega_1$ and $\Omega_2$ in Equation
(\ref{3.18}) gives $\frac{^Nd\hat {\bf n}_h}{dt} = {\bf 0}$, which explains
that $\hat {\bf n}_h$ is an inertially fixed unit vector thus, also making $Q$
an inertial frame. By extension, it can be inferred that the angular momentum of
a variable mass system is also an inertially fixed vector as it is directed
along $\hat {\bf n}_h$.

Through this write-up it has been shown that the angular momentum of a variable
mass system is inertially fixed despite its variable magnitude. As mentioned in
the earlier portions of this text, this result can serve as the foundation for
analytical and geometric examinations of the rotational motions of variable mass
systems in a manner similar to that seen in the literature on rigid bodies. For
example, one may attempt to perform a stability analysis of rotating variable
mass systems. Further, as the theory developed here is applicable to flexible
systems it can be useful in studies on engineering systems such as rockets, and
balloons. This result can also find application in navigation and control of
underwater vehicles which are modeled using a control volume approach.

\end{document}